\documentclass{PoS}
\usepackage{cleveref}
\usepackage{wrapfig}
\usepackage{amsmath}
\usepackage{graphicx}
\usepackage[caption=false]{subfig}
\usepackage[left]{lineno}
\usepackage{enumitem}
\usepackage{float}

\title{IceTop as veto for IceCube: results}

\ShortTitle{IceTop as veto for IceCube: results}

\author{
The IceCube Collaboration\footnote{For collaboration list, see PoS(ICRC2019) 1177.}\\
{\itshape \href{http://icecube.wisc.edu/collaboration/authors/icrc19_icecube}{http://icecube.wisc.edu/collaboration/authors/icrc19\_icecube}}\\
E-mail: \email{delia.tosi@icecube.wisc.edu, hpandya@icecube.wisc.edu}
}

\abstract{
The IceCube Neutrino Observatory features both a kilometer-cubed detector between 1.45 and 2.45 km depth and an array of ice-filled tanks, called IceTop, located at the surface. The presence of both detectors at the same location allows for detailed studies of cosmic rays and their muon content in ice, while the lack of signals in the surface detectors can be used to identify muon tracks in the deep detector as neutrino candidates and to determine the veto efficiency of IceTop. While the solid angle coverage of the current detectors is limited, this has interesting implications for the design of a larger surface array.  In this contribution, we present the results from this study applied to 5 years of data. We find a few interesting neutrino candidate events that pass the cuts designed to veto cosmic rays.  Thorough simulations are necessary to establish the likelihood for these events to be astrophysical neutrinos or rare cosmic rays. \\

\vspace{4mm}
{\bfseries Corresponding authors:}
\speaker{Delia Tosi}$^{1}$, Hershal Pandya$^{2}$\\
{$^{1}$ \itshape Dept. of Physics and Wisconsin IceCube Particle Astrophysics Center, University of Wisconsin, Madison, WI 53706, USA}\\
{$^{2}$ \itshape Bartol Research Institute and Dept. of Physics and Astronomy, University of Delaware, Newark, DE 19716, USA}

}

\FullConference{36th International Cosmic Ray Conference -ICRC2019-\\
		July 24th - August 1st, 2019\\
		Madison, WI, U.S.A.}

\begin{document}
\setlength{\belowdisplayskip}{5pt} \setlength{\belowdisplayshortskip}{5pt}
\setlength{\abovedisplayskip}{0pt} \setlength{\abovedisplayshortskip}{0pt}
\section{Introduction}\label{sec:intro}
IceCube is a neutrino detector located at the geographic South Pole \cite{detector}. More than five thousands photosensors (Digital Optical Modules or DOMs) are installed in a cubic-kilometer ice volume between depths of 1450 m and 2450 m. The sensors feature photomultiplier tubes that detect the Cherenkov radiation emitted by charged particles traveling through the ice volume. Above IceCube lies IceTop, a surface detector which uses the same detection technology as IceCube, with photosensors embedded in 162 frozen water tanks installed in pairs in proximity of each string \cite{IceTopPaper}.

\begin{wrapfigure}{r}{0.45\textwidth}
	\vspace{-20pt}
	\begin{center}
		\includegraphics[width=0.42\textwidth]{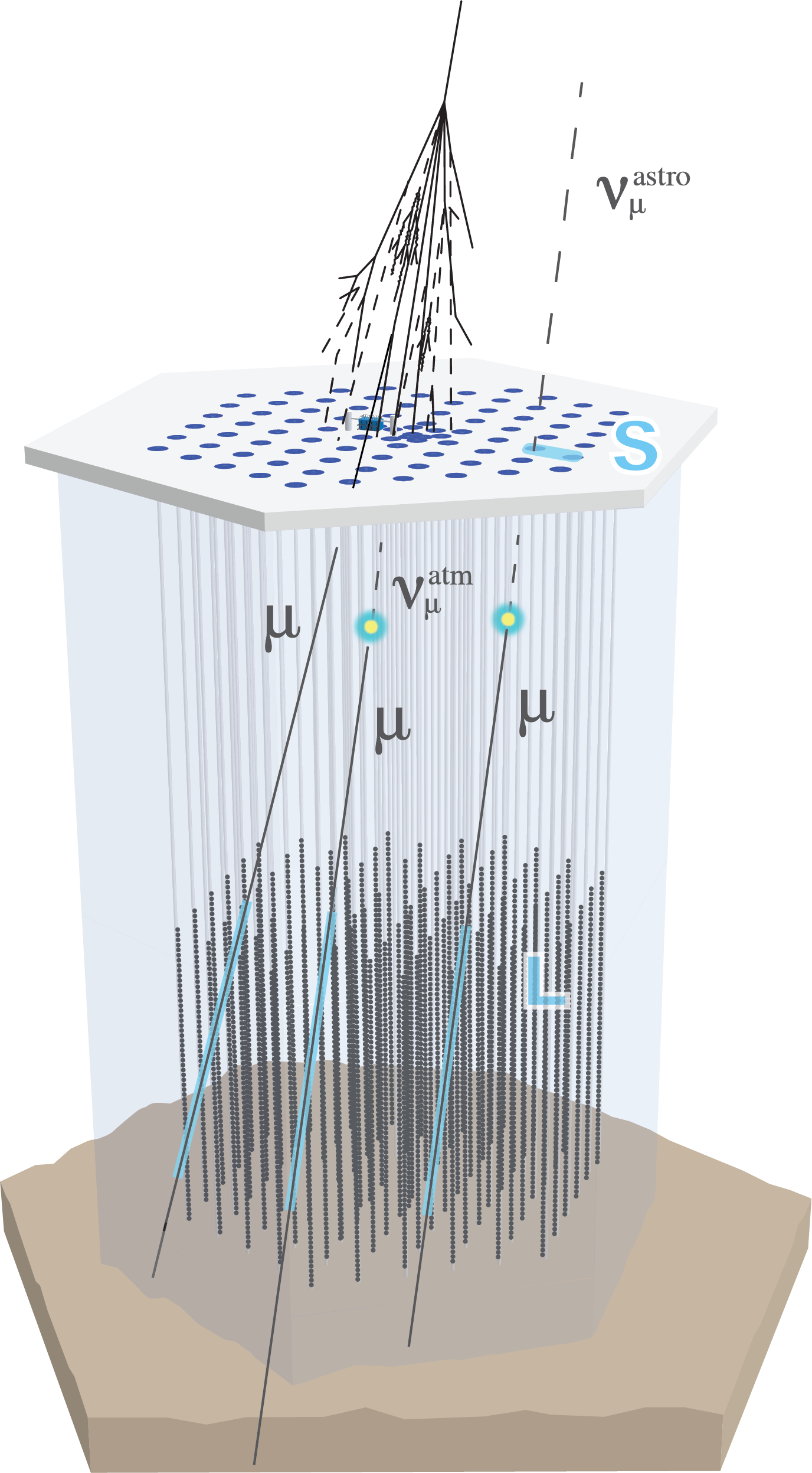}
		\vspace{-10pt}
		\caption{Illustration of the containment cuts used to select events for this analysis: track length \mbox{L$\geq$800\,m} and side distance \mbox{S$\geq$60\,m}. Additional quality cuts are applied as described in the text.}
		\label{fig:sketch}
	\end{center}
	\vspace{-20pt}
\end{wrapfigure}

Muon tracks are the most important event topology in IceCube data, as their good pointing resolution makes them ideal candidate for real-time neutrino alerts and point sources searches \cite{MultiMessenger}. 
The sought-after tracks are those from single muons produced by astrophysical muon neutrinos undergoing a charged current interaction. The interaction vertex depth (if inside the instrumented volume) and the direction and energy of the muon provide an indication of the likelihood of the track to have generated by an astrophysical $\nu_{\mu}$ rather than by an atmospheric neutrino \cite{HESE}. 

The vast majority of tracks detected by IceCube are however muons or muon bundles produced by cosmic-ray showers that penetrate to the instrumented in-ice volume. 
These muons constitute the largest background to astrophysical neutrino searches from the Southern Hemisphere. Currently no other method exists to identify neutrino-induced tracks from the Southern Hemisphere whose interaction vertex lies outside the instrumented volume. 

IceTop data can be used to `veto' the muon tracks produced by cosmic-ray showers, since the sensors in the tanks will detect the shower particles traveling in the tanks (hits). Thus a cosmic-ray-induced muon track will be characterized by hits in the surface detector while a neutrino-induced one by the lack thereof. While many tracks in the selection will also trigger IceTop, the analysis aims at vetoing also muon-related showers which may generate hits sparse enough that they do not generate a trigger in IceTop.

The goal of this analysis is to identify the veto capabilities of IceTop by looking at the correlation between IceTop hits and muon tracks in 5 years of data (April~2012 to May~2017). This work completes what was presented in \cite{ICRC2015} and \cite{ICRC2017} and prepares for the addition of interesting events found using this method to the real time neutrino alert stream. 

\section{Data Selection}
The dataset for this analysis consists of muon tracks selected using only the in-ice detector. A good angular and arrival-time reconstruction is necessary to correlate with correct IceTop hits, and hence we require that each in-ice event must satisfy the following conditions:
\begin{itemize}[noitemsep,topsep=0pt]
    \item The sum of the charges recorded by all the DOMs (excluding the brightest DOM) must exceed $10^3$ p.e. (photoelectrons).
    \item The zenith angle, $\theta$, must satisfy $\cos(\theta) \geq 0$, to ensure selection of down-going tracks.
    \item The intersection point of the track on the surface must be at least 60~m inside the IceTop boundary.
    \item The point on the track, nearest to the center of the in-ice detector, must be located within the inner 85\% volume of the detector.
    \item The muon track length must be at least 800~m.
    \item The angle between the track directions determined by the main reconstruction algorithm and another independent algorithm must be less than 15$^\mathrm{o}$.
    \item The value of the fit quality parameter for the main reconstruction, RLogL (analogous to $\chi^2$), must be less than 8.
\end{itemize}
The main reconstruction algorithm used for this analysis takes into account the time of the first pulse but also the total charge detected by the DOM and uses spline fits to the ice properties to calculate the most likely direction of a track \cite{DireReco}.  
The cuts described above were checked against simulations of air showers (produced with \textsc{CORSIKA}) and neutrinos (produced with \textsc{NeutrinoGenerator} \cite{NuGen}). A sketch depicting some of the selection cuts, together with the analysis idea, is shown in Fig.\,\ref{fig:sketch}.

\begin{wrapfigure}{r}{0.45\textwidth}
    \vspace{-20pt}
	\begin{center}
    \includegraphics[width=.45\textwidth]{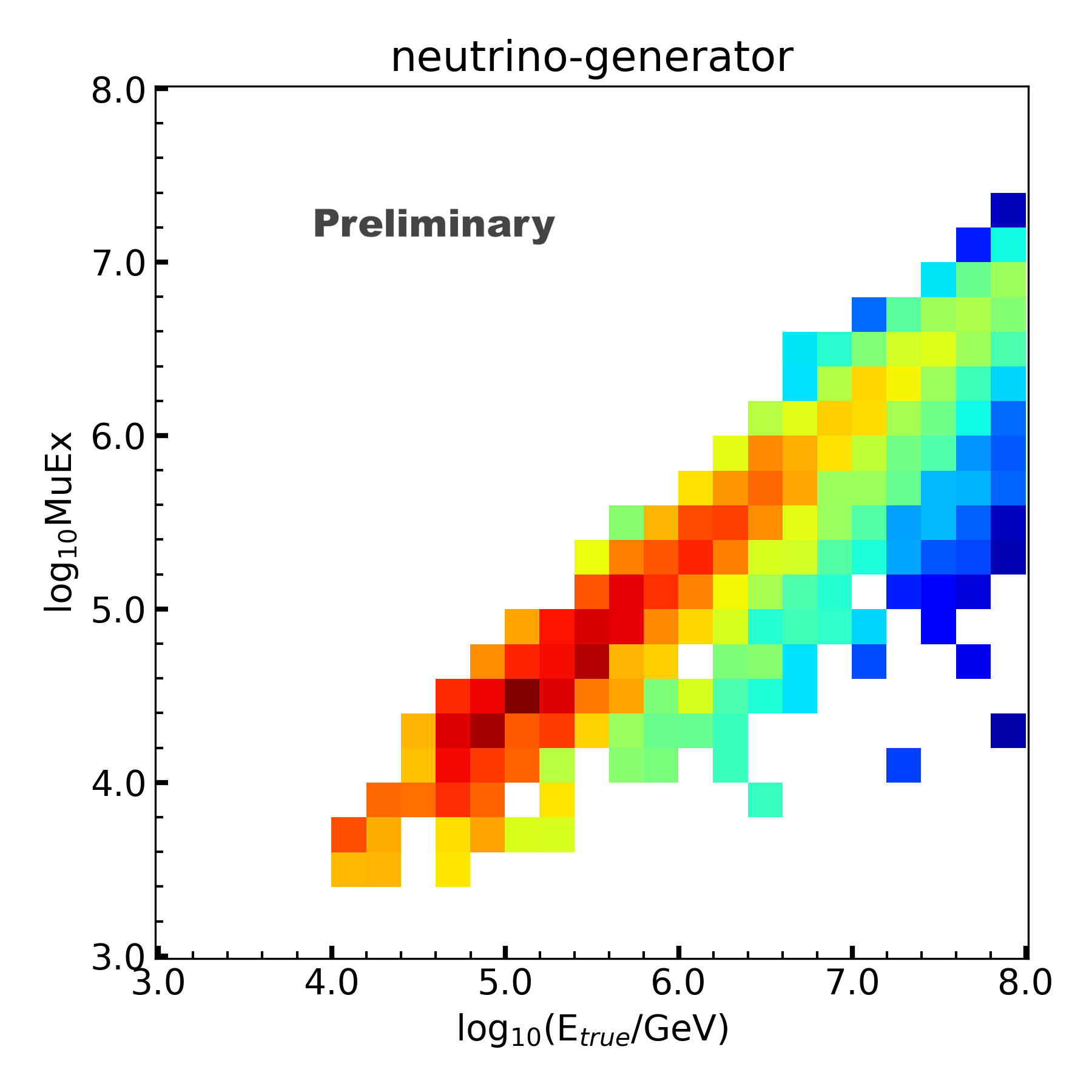}
    \vspace{-30pt}
    \caption{Relation between MuEx, the muon energy proxy used in this analysis and the primary neutrino energy for an astrophysical flux.}
    \label{fig:energy_cal}
    \end{center}
	\vspace{-18pt}
\end{wrapfigure}

The energy proxy used in this analysis, called MuEx, is determined by fitting the expected number of photons via an analytic template which scales with the energy of the muon. This energy estimator accounts for energy losses outside the detector and it is therefore more accurate than a simple sum of the DOM charges \cite{EnergyReco}. A minimum energy of $
\log_{10}(\mathrm{MuEx}) = 3.0 $ is required for a track to be included in the current analysis.  A calibration for the neutrino energy as a function of the energy proxy is shown in Fig.~\ref{fig:energy_cal} using a two dimensional histogram re-weighted to the best fit astrophysical neutrino spectrum with spectral index -2.13$\pm$0.13~\cite{Diffuse6years}.

To apply the IceTop based event discrimination method described in Section~\ref{sec:method} it is necessary to have datasets of cosmic-ray and neutrino-like events. The observed muon track events and the accompanying IceTop hits form the cosmic-ray dataset since the fraction of neutrino-like events is expected to be negligible in data. We generate a neutrino-like dataset by replacing the IceTop signals for observed muon tracks with background IceTop hits. The background IceTop hits are usually from small showers or stray particle hits that fail to satisfy IceTop's air shower trigger condition. We extract these background hits from IceTop recordings made by the IceCube data acquisition system via unbiased periodic forced triggers. With this method the analysis is completely data-driven with no dependence on simulations.
\section{Method}\label{sec:method}
\begin{figure}[H]
    \centering
    \captionsetup[subfloat]{farskip=1pt,captionskip=-2pt}
    \subfloat[Cosmic-ray PDF ($H_\mathrm{CR}$)]
    {\includegraphics[width=0.49\textwidth]{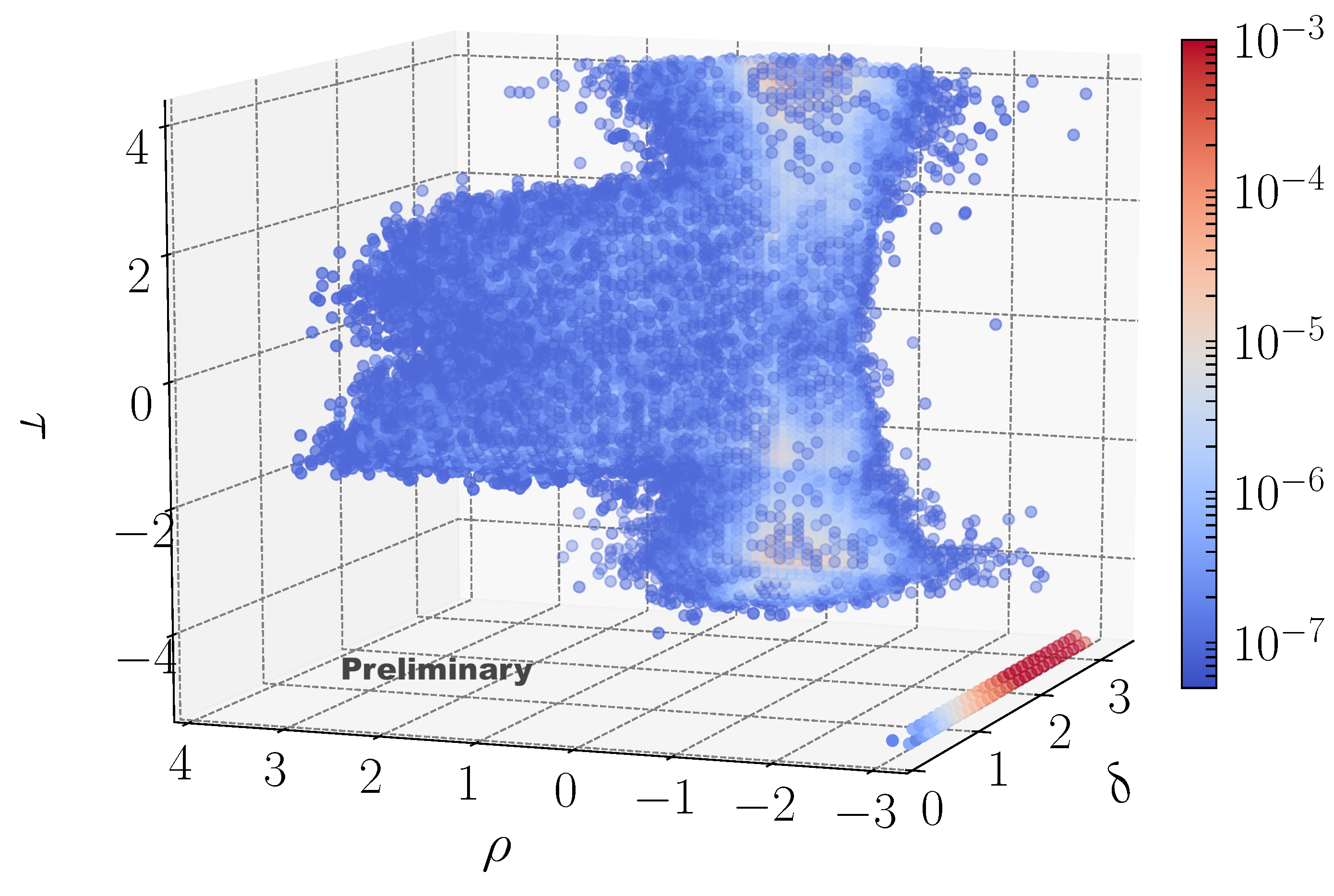}\label{fig:PDF_CR}}
    \subfloat[Neutrino-like PDF ($H_\nu$)]
    {\includegraphics[width=0.49\textwidth]{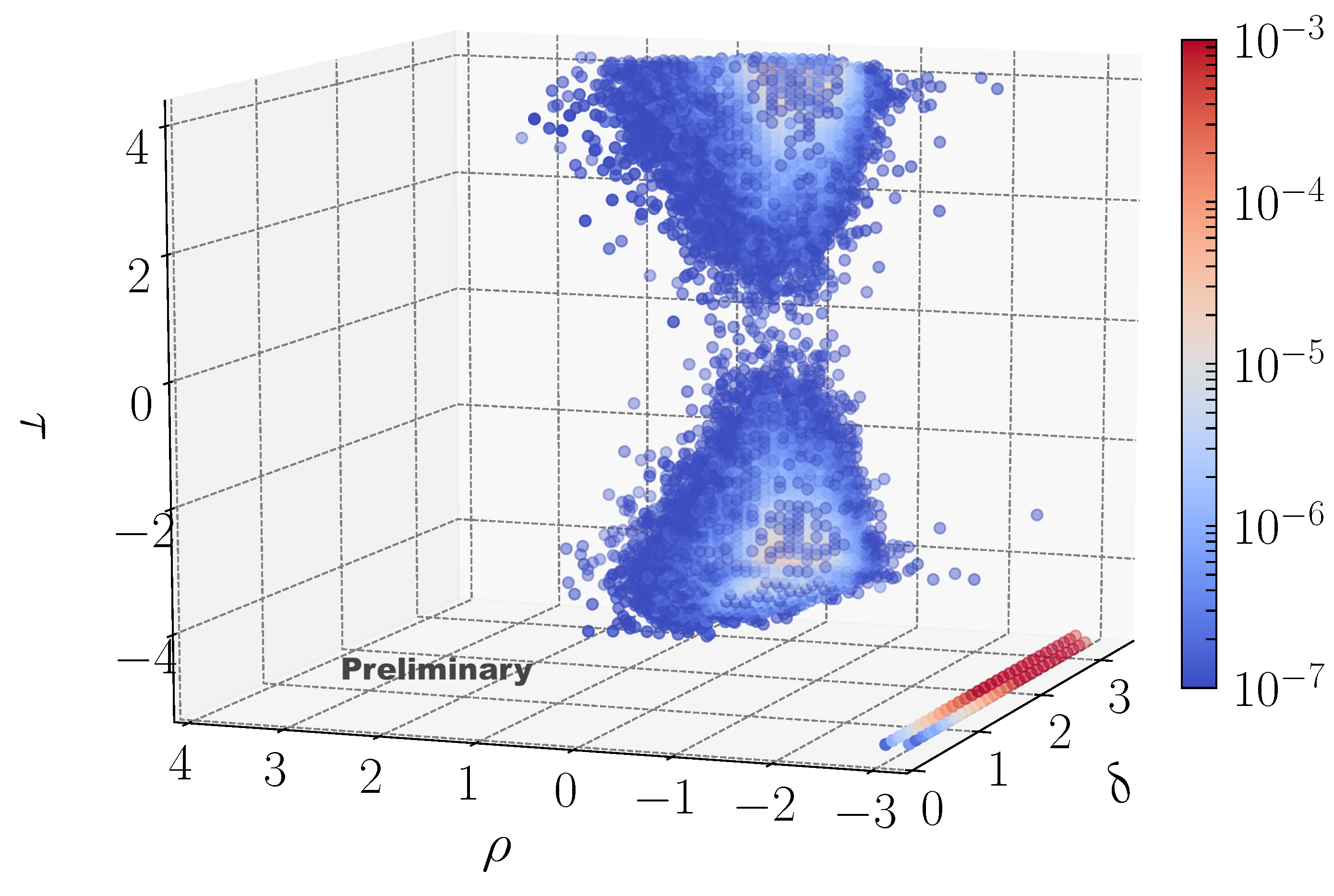}\label{fig:PDF_Nu}}
    \caption{Figs.(a) and (b) show the PDFs for cosmic-ray and neutrino-like events respectively, for \mbox{$4.2 \leq \log_{10}(\mathrm{MuEx}) < 4.4$} and \mbox{$0.96 \leq \cos(\theta) < 0.98$} for the data-year 2012. The region of the PDFs near $\rho = -3$ and $\tau = -5$ contains the un-hit tanks and non-functional tanks that were assigned a fixed value for charge and time outside the normal range of values.}\label{fig:onebin}
\end{figure}

\begin{wrapfigure}[16]{r}{0.45\textwidth}
    \vspace{-30pt}
	\begin{center}
    \includegraphics[width=.45\textwidth]{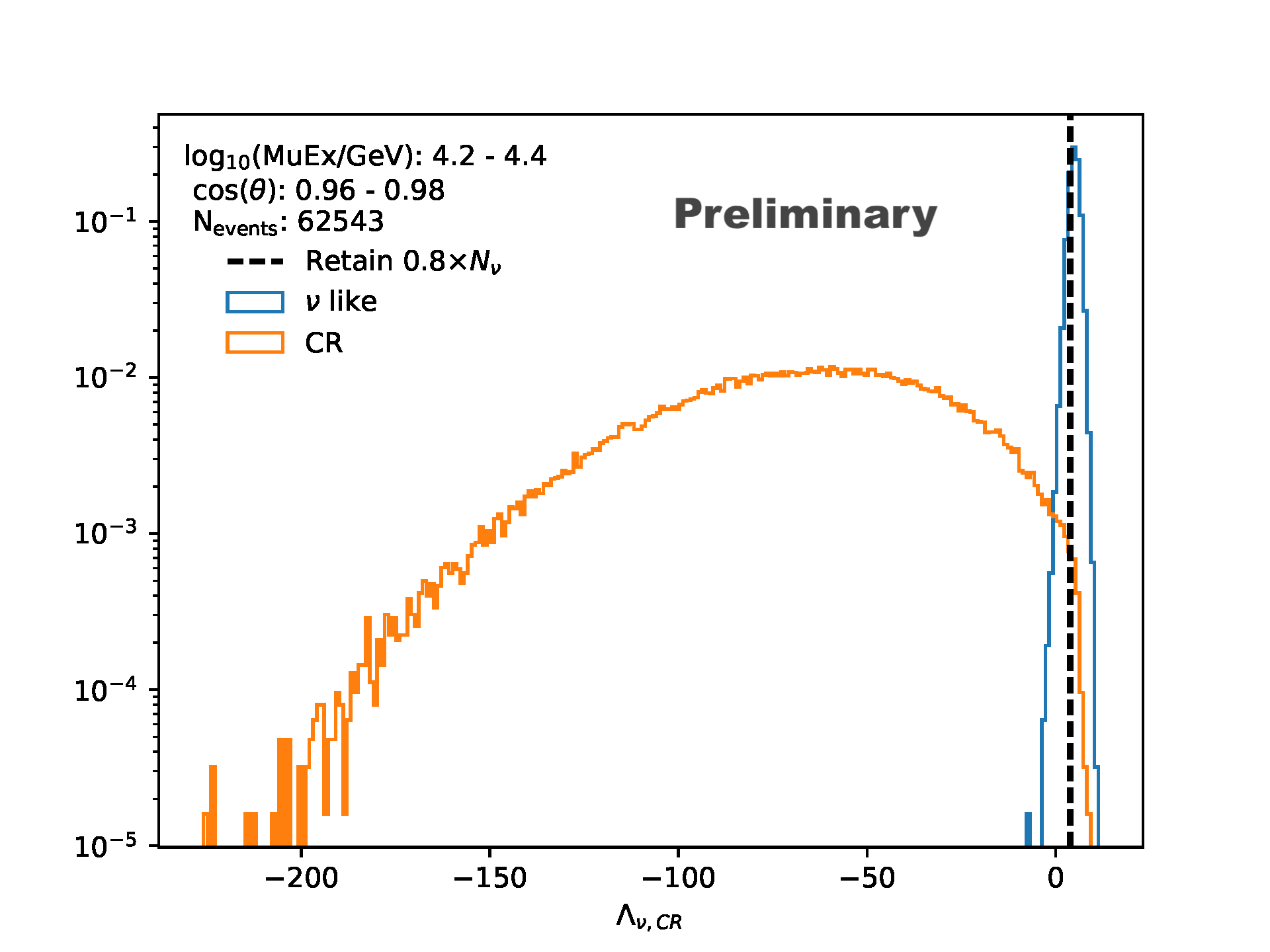}
    \vspace{-25pt}
    \caption{The distribution of IceTop log-likelihood ratios calculated for cosmic-ray (orange) and neutrino-like (blue) events in the bin shown in Fig.\,\ref{fig:onebin}. A dotted line shows the value of $\Lambda_{\mathrm{cut}}$ chosen to retain 80\% of neutrino-like events.}
    \label{fig:LLHR}
    \end{center}
	\vspace{-40pt}
\end{wrapfigure}

We employ the IceTop log-likelihood ratio method originally presented in \cite{ICRC2017} and \cite{Pandya2017} to distinguish shower-related hits in IceTop from hits uncorrelated with the muon track. Before doing the analysis, all the events are binned into energy ($\log_{10}(\mathrm{MuEx})$) and zenith ($\cos(\theta)$) bins. For each event, we extend the muon track to the surface. If a hit has been recorded in a tank within the event readout window, we record the charge (Q) detected in units of vertical equivalent muon (VEM), the residual time ($t_{\mathrm{res}}$) compared to the expected shower front arrival time (using a data-derived model for the shower front curvature), and the perpendicular distance of the tank from the shower axis (d). From charge, residual time and distance we then define three variables using the following coordinate transformation:
\begin{equation}
    \begin{aligned}
    \rho = \log_{10}(\mathrm{Q/[VEM]}) \\
    \tau = \text{sign}(t_{\mathrm{res}}/\mathrm{[ns]})\; \log_{10}( |t_{\mathrm{res}}/\mathrm{[ns]}| + 1 ) \\
    \delta = \log_{10}( \mathrm{d}/\mathrm{[m]} + 1) .
    \end{aligned}
    \label{eq:rhotaudelta}
\end{equation}

Using the observables defined in Equation~\ref{eq:rhotaudelta} we construct three-dimensional probability density functions (PDFs) for both the cosmic-ray sample ($H_\mathrm{CR}$) and the neutrino-like sample ($H_\nu$) using all the events in each dataset. The PDFs are shown in the Figs.~\ref{fig:PDF_CR} and ~\ref{fig:PDF_Nu} for one example $\log_{10}(\mathrm{MuEx})$ and $\cos(\theta)$ bin. Tanks which do not have any hits, and tanks which were not taking data properly during the run also contribute to the PDF with pre-assigned values for the time and the charge, outside the standard range of values. The log-likelihood ratio for a given event $x_j$ is then calculated as:%
\begin{equation}
\Lambda_{\nu, CR} (x_j) = \log_{10}\left(\frac{\prod_{\mathrm{i}=1}^{162} \\ \mathrm{P}\left(\rho_i, \  \tau_i, \ \delta_i \ | \ H_\nu \right)}{\prod_{\mathrm{i}=1}^{162} \\ \mathrm{P}\left(\rho_i, \ \tau_i, \ \delta_i \ | \ H_{CR} \right)}\right)
\end{equation}%
where $\mathrm{P}\left(\rho_i, \ \tau_i, \ \delta_i \ | \ H \right)$ is the probability for observing a tank with such a charge, time and distance under the hypothesis $H$. Before calculating $\Lambda_{\nu, CR}$ for an event $x_j$, the contribution of the event itself is subtracted out of the corresponding hypothesis ($H_\nu$ or $H_\mathrm{CR}$) to avoid over-fitting in bins with low statistics. If for a given ($\rho_i$, $\tau_i$, $\delta_i$), the value of $P_i$ is not available then an extrapolation is done to obtain it from the nearest populated bins in the PDF.


The PDFs and the $\Lambda_{\nu, CR}$ distributions are shown for the cosmic-ray and for the neutrino-like events in Fig.\,\ref{fig:onebin} and Fig.\,\ref{fig:LLHR} for one example zenith and energy bin. We fix the cut value ($\Lambda_{\mathrm{cut}}$) in each bin so as to retain 80\% of the neutrino-like sample but the retention percentage is increased to 99.9\% for bins with $\log_{10}(\mathrm{MuEx}) \geq$ 5.2. The events with $\Lambda_{\nu, CR} < \Lambda_{\mathrm{cut}} $ are labeled as cosmic-ray muons and vetoed. 

\section{Results}\label{sec:results}
\begin{figure}[H]
    \centering
    \captionsetup[subfloat]{farskip=1pt,captionskip=-2pt}
    \subfloat[2012-2013]
    {\includegraphics[width=.5\textwidth]{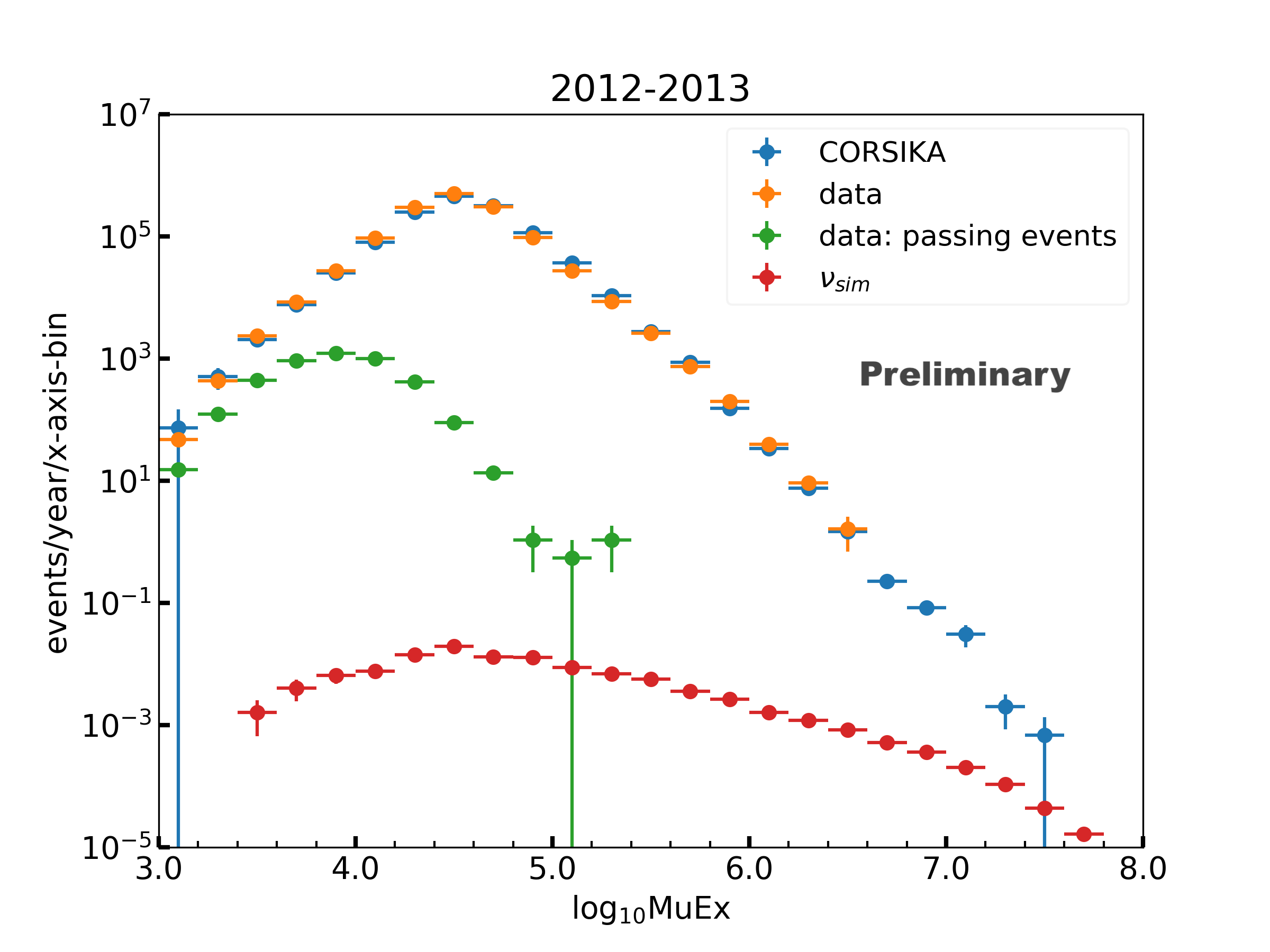}\label{fig:counts_muex_1213}}
    \hspace{-5pt}
    \subfloat[2014-2015-2016]
    {\includegraphics[width=.5\textwidth]{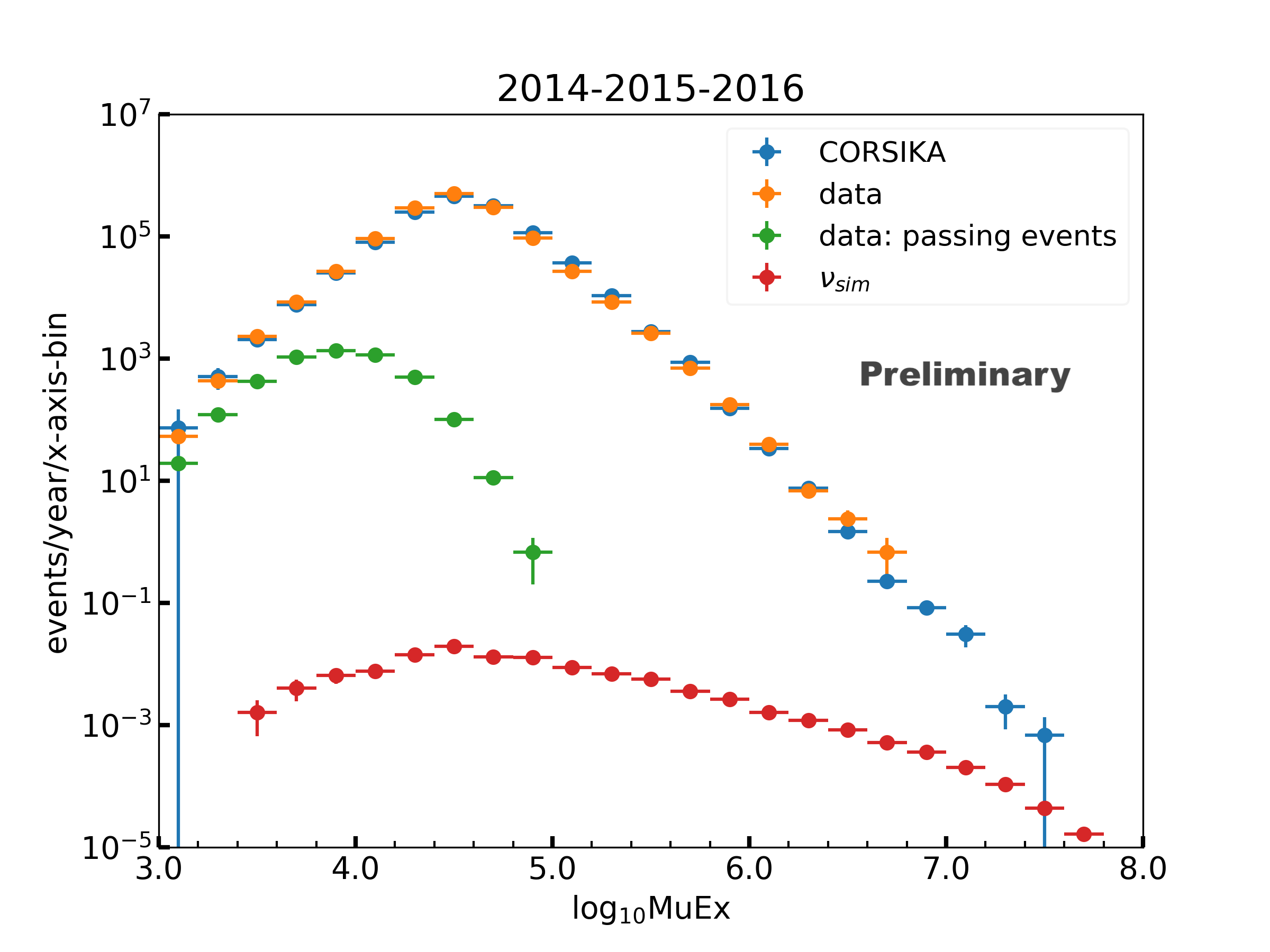} \label{fig:counts_muex_1415}}
    \hspace{-5pt}
    \caption{Number of events per year and bin as a function of muon energy proxy for CORSIKA, simulated neutrinos, and data (2012-2013 and 2014-2015-2016) before and after the veto cuts.}\label{fig:counts_muex}
\end{figure}

We analyzed each year separately and verified consistency of the results between the different years. Due to an error in the data acquisition system, 11~tanks in the 2012-2013 data were not active as normal during in ice triggers - these tanks have been excluded from the analysis for these two years. We therefore merge the data collected in the 5~years in two separate samples. Since the data-taking switchover from one year to the following happens sometime around May (with initial test runs), the sample designated as "2012-2013" includes approximately 674.45 days between April 26$^\mathrm{th}$, 2012 and May 6$^\mathrm{th}$, 2014. The sample designated as "2014-2015-2016" includes approximately 1067.35 days between April 10$^\mathrm{th}$ 2014 and May 18$^\mathrm{th}$, 2017. The PDFs and the $\Lambda_{\mathrm{cut}}$ are calculated for each of these two samples. The final results of this analysis are summarized in Figs.~\ref{fig:counts_muex_1213} and \ref{fig:counts_muex_1415}. Each of these figures shows, as a function of $\log_{10}(\mathrm{MuEx})$, the number of events in the data before and after the IceTop log-likelihood ratio cut, the number of simulated cosmic-ray (CORSIKA) events weighted by H3a~\cite{h3a} spectrum model, and the number of simulated neutrino events weighted by the neutrino astrophysical flux measured by IceCube $\phi_{\nu+\bar{\nu}}$ = $\left(0.90^{+0.30}_{-0.27}\right)\times$ 10$^{-18}$GeV$^{-1}$cm$^{-2}$s$^{-1}$sr$^{-1}$ $\cdot$(E$_{\nu}$/100\,TeV)$^{\gamma}$ with $\gamma\,=\,-2.13\pm0.13~$~\cite{Diffuse6years}. 

\begin{figure}[H]
\centering
\includegraphics[width=0.9\textwidth]{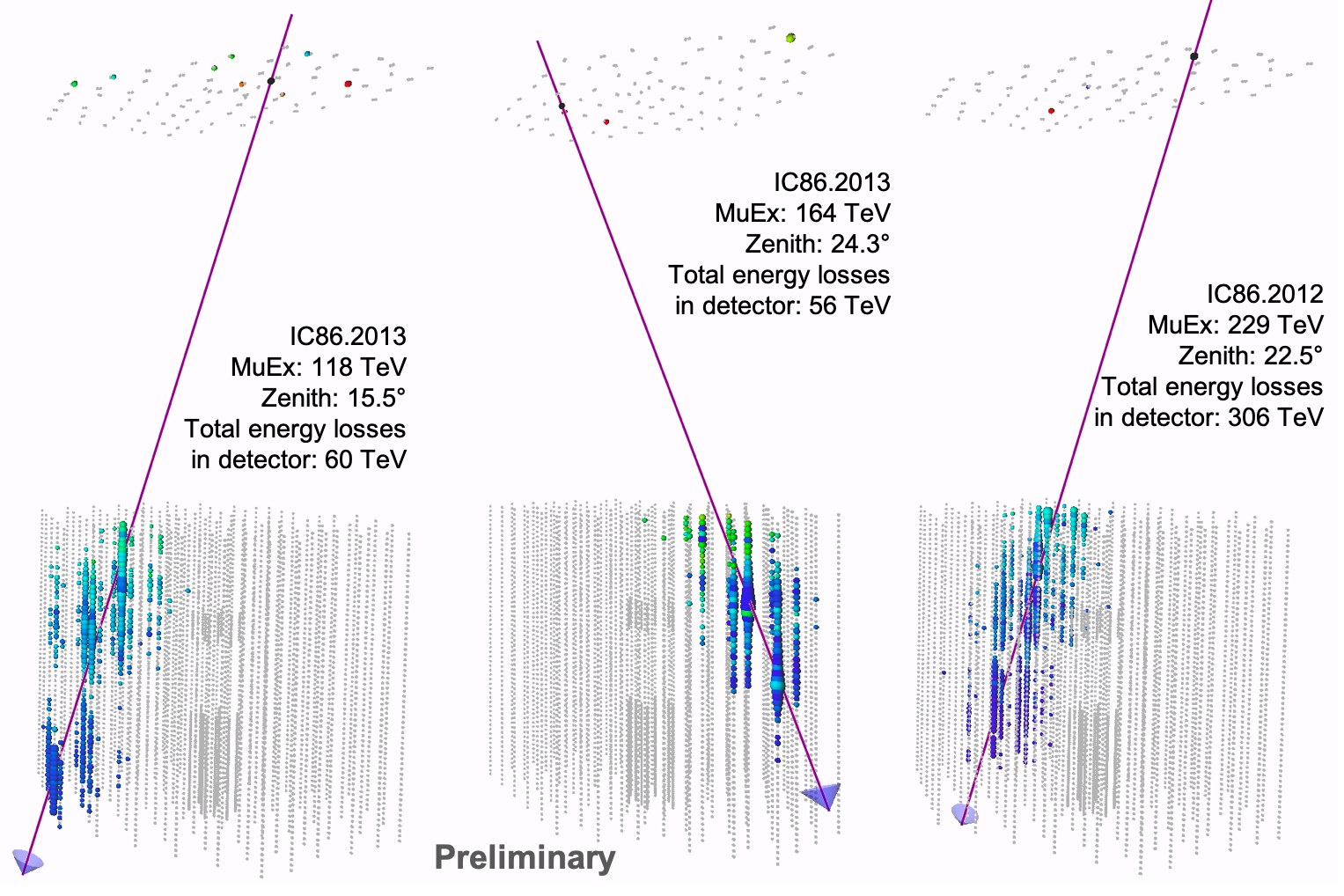}\caption{Detector view of interesting passing events with  $\log_{10}(\mathrm{MuEx})\geq5.0$. The detector has been oriented differently in each frame to optimize the event view.}
\label{fig:events}
\end{figure}

Especially of interest are the muon tracks of the highest energies that pass the IceTop veto. The statistical significance of these passing muon events cannot be evaluated in absence of thorough cosmic-ray simulations with a complete in-ice and IceTop response. However, the highest energy muons that pass the IceTop veto are candidate astrophysical neutrinos.  In the 2012-13 sample, 5~events out of 249694 pass the cuts for $\log_{10}(\mathrm{MuEx})\geq 4.8$ out of which 3~passing events have $\log_{10}(\mathrm{MuEx})\geq 5.0$. In the 2014-15-16 sample, 2~events over 387576 pass the cuts for $\log_{10}(\mathrm{MuEx})\geq 4.8$ and no passing event has $\log_{10}(\mathrm{MuEx})\geq 5.0$. Passing events with $\log_{10}(\mathrm{MuEx})\geq 5.0$ are displayed in the Fig.~\ref{fig:events} along with their IceTop hits. Each DOM (IceTop and in-ice) that was hit during this trigger readout window has been coloured according to their time stamp - with red being early and blue being late.

Another indicator useful to determine if a track is more neutrino-like or cosmic ray-like is stochasticity. Muons experience a constant ionization energy loss as they pass through the ice and large radiative losses stochastically due to Bremsstrahlung. A muon neutrino interaction produces a single muon whereas cosmic-ray showers most often produce a bundle of high energy muons collimated along the shower axis. Energy losses along the track will be more stochastic for a single muon, whereas bundles will appear to lose energy more smoothly. 
\begin{figure}[H]
\centering
\includegraphics[width=1\textwidth]{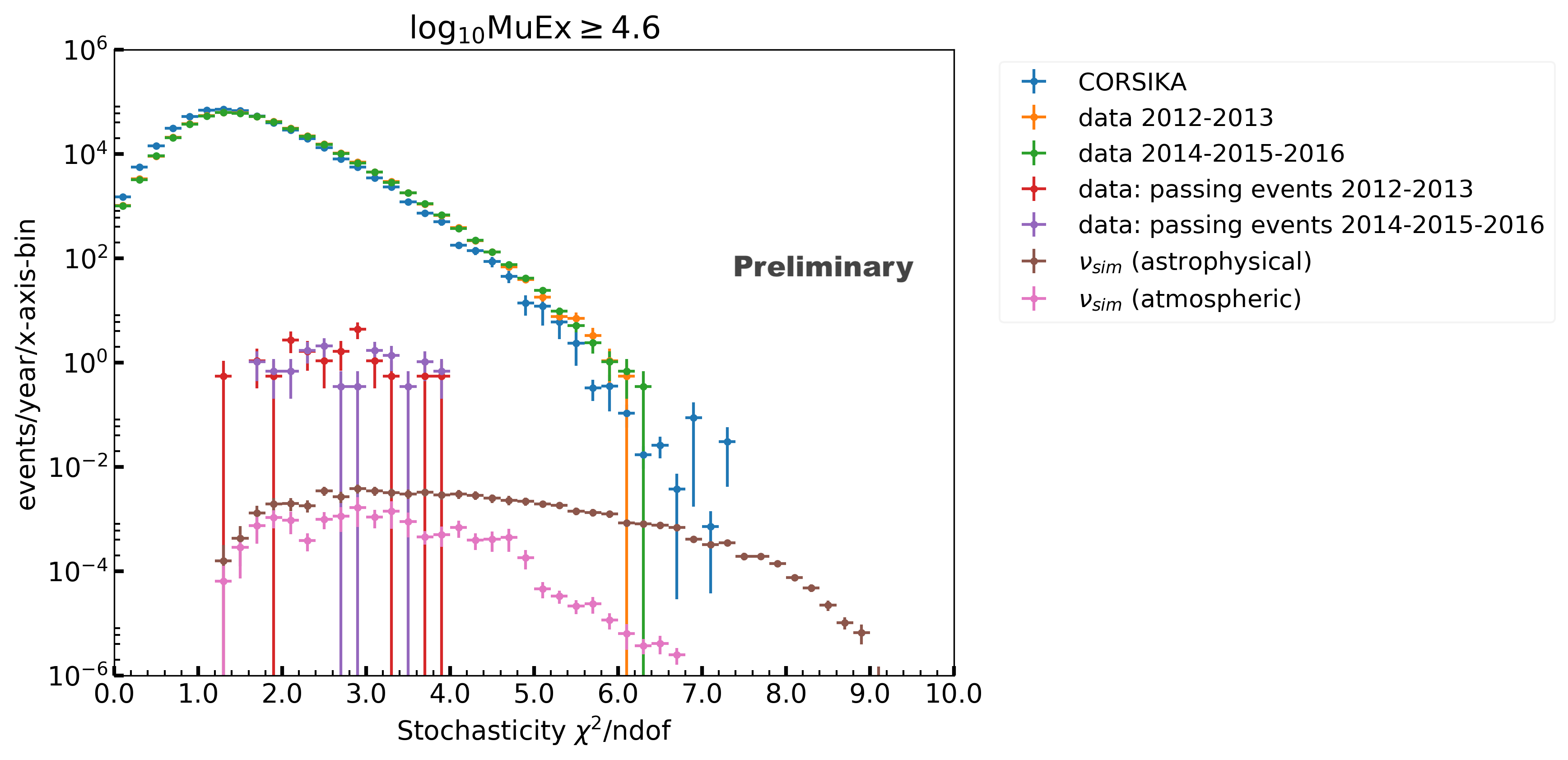}
\vspace{-20pt}
\caption{Stochasticity distribution for CORSIKA events, data sample, passing events in the sample, neutrino simulation weighted to atmospheric neutrino spectrum and neutrino simulation weighted to astrophysical spectrum. The parameter on the x-axis is used as an indicator for stochastic losses along the muon/muon bundle track. Single muons (such as those produced by neutrinos) have a distribution with average higher values than that typical of muon bundles. For a detailed explanation of this parameter see text.}
\label{fig:stoch}
\end{figure}
\begin{wrapfigure}[15]{r}{0.45\textwidth}
    \vspace{-20pt}
	\begin{center}
    \includegraphics[width=.45\textwidth]{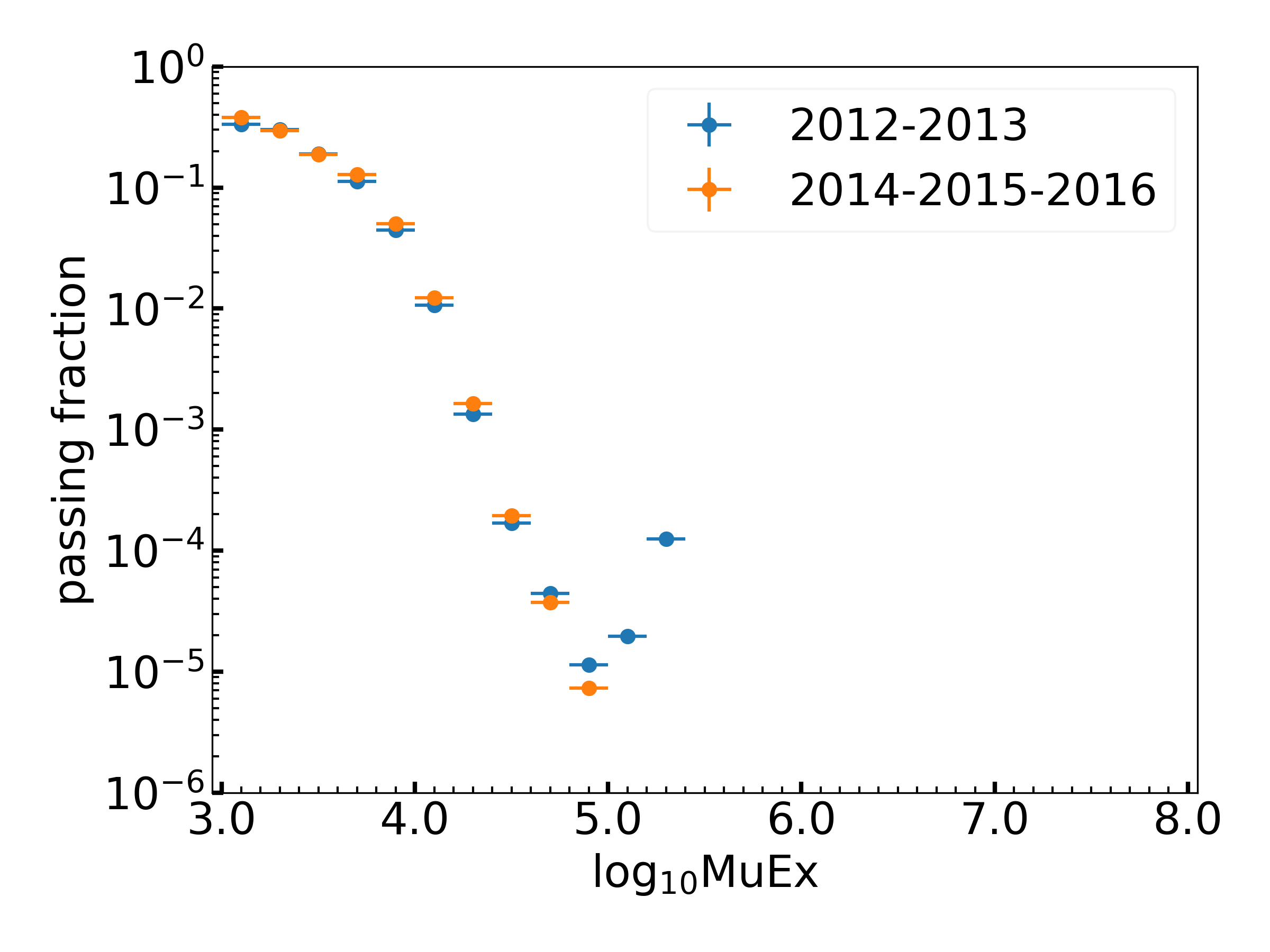}
    \vspace{-25pt}
    \caption{Passing fraction as a function of muon energy proxy for 2012-2013 and 2014-2015-2016, calculated as ratio of passing to total events from Fig.\ref{fig:counts_muex}.}
    \label{fig:passing_fraction}
    \end{center}
	\vspace{-40pt}
\end{wrapfigure}
For each track, we fit the energy losses along the track length to a linear function. The reduced-chi-square ($\chi^2$/ndof) from the fit is a measure of muon multiplicity, with higher reduced-chi-square values corresponding to larger stochasticity i.e. to single muons. The expected distribution for simulated muon neutrino events (both atmospheric and astrophysical) and for CORSIKA events is shown in Fig.~\ref{fig:stoch}. The stochasticity of passing events has also been plotted on the same plot. While this analysis has not been developed with focus on this parameter, it appears that the stochasticity distribution of passing events is compatible with both muon bundles and neutrino-induced single muons.   

Assuming that all the passing events are cosmic rays that sneak through the veto, we can use the counts shown in Fig.\,\ref{fig:counts_muex} to calculate the veto passing fraction, shown in Fig.\,\ref{fig:passing_fraction} as a function of energy proxy. IceTop achieves a reduction of $2\times10^{-5}$ to $5\times10^{-6}$ in the atmospheric background for down-going muon neutrino sample for a minimum neutrino energy of $\sim$100~TeV.

\section{Conclusions}

The method presented here rejects in-ice muons related to small showers, by comparing the IceTop footprint (charge, time and distance) of recorded hits to that of typical showers. We have shown that IceTop can achieve a reduction of $2\times10^{-5}$ to $5\times10^{-6}$ in the atmospheric background for down-going muon neutrino sample for a minimum neutrino energy of $\sim$100~TeV. 
Along with down-going cosmic-ray muons, this method also vetoes the muon tracks from muon neutrinos of atmospheric origin using the footprint of accompanying shower at IceTop surface. The rejection efficiency of atmospheric muon neutrinos  will be studied with detailed simulations.

The passing events found as a result of this analysis are candidate astrophysical neutrinos from the Southern Hemisphere. The selection can be further tuned to retain a target number of interesting events per year, and could be integrated in the real neutrino alert program in the future. The version of the IceTop log-likelihood ratio developed for this analysis can be easily applied to the simulated future extensions of the surface array \cite{Gen2SurfaceVeto}. This will inform the best possible detector design by providing a realistic veto efficiency achievable by various array configurations.



\bibliographystyle{ICRC}


\end{document}